\documentclass
[10pt,letterpaper,bibnotes,notitlepage,final,balancelastpage,superscriptaddress,twocolumn]{revtex4}%
\usepackage{amssymb}
\usepackage{amsmath}
\usepackage{amsfonts}
\usepackage{graphicx}
\usepackage{bm}
\usepackage{float}
\usepackage[bookmarks=false,pdfstartview=FitH,hyperindex=true, colorlinks,
linkcolor=blue, citecolor=blue]{hyperref}
\usepackage[all]{hypcap}%
\providecommand{\U}[1]{\protect\rule{.1in}{.1in}}
\begin{document}

\title{Chiral Spin Liquid in Correlated Topological Insulator}
\author{Jing He}
\affiliation{Department of Physics, Beijing Normal University, Beijing, 100875 P. R.
China }
\author{Su-Peng Kou}
\thanks{Corresponding author}
\email{spkou@bnu.edu.cn}
\affiliation{Department of Physics, Beijing Normal University, Beijing, 100875 P. R.
China }
\author{Ying Liang}
\affiliation{Department of Physics, Beijing Normal University, Beijing, 100875 P. R.
China }
\author{Shiping Feng}
\affiliation{Department of Physics, Beijing Normal University, Beijing, 100875 P. R.
China }

\begin{abstract}
In this paper, we investigate the topological Hubbard model - the spinful
Haldane model with on-site interaction on honeycomb lattice with spin
rotation symmetry by using slave-rotor approach and find that chiral spin
liquid exists in such a correlated electron system of the intermediate
coupling region. By considering the anyon nature of excitations, chiral spin
liquid may be the ground state of the topological Hubbard model. The low
energy physics is basically determined by its Chern-Simons gauge theory.

\end{abstract}

\maketitle

The Fermi liquid based view of the electronic properties has been very
successful as a basis for understanding the physics of conventional solids
including metals and (band) insulators. For the band insulators, due to the
energy gap, the charge degree of freedoms are frozen. If there exist
spontaneous spin rotation symmetry breaking, the elementary excitations are
the gapless spin wave and the gapped quasi-particle (an electron or a hole)
that carry both spin and charge quantum numbers. However, in some special
insulators, the elementary excitations with fractional quantum numbers of an
electron may exist. People call them quantum spin liquid states\cite%
{pwa_87,wxg,7}. \ There exist different types of ansatz of spin liquid: $%
\mathrm{Z}_{2}$, $\mathrm{U(1)}$, $\mathrm{SU(2)}$ and $\mathrm{SU(2)\times
SU(2)}$\cite{wxg,7}. These different spin liquid states have the exactly the
same global symmetry, as conflicts to Landau's theory, in which two states
with the same symmetry belong to the same phase. In particular, there exist
quantum spin liquid states breaking time reversal symmetry, of which the
elementary excitations are anyons with fractional statistics. People call
them \emph{chiral spin liquid (CSL)}\cite{chiral,7}.\emph{\ }There are two
types of CSLs - the abelian CSL with abelian anyonic excitations and
non-Abelian CSL with non-Abelian anyons.

Recently, nonAbelian CSL state has been predicted in the Kitaev model on
honeycomb lattice or in its generalizations\cite{k2}. On the contrary,
although Abelian CSL has been proposed much earlier than non-Abelian CSL,
till now people don't know any types of model with the (abelian) chiral spin
liquid as the ground state. Then one issue here is\emph{\ may people realize
chiral spin liquid in certain many-body systems}? To answer above question
we study the quantum properties of the so-called \emph{topological Hubbard
model} on honeycomb lattice with spin rotation symmetry by using slave-rotor
approach and propose that chiral spin liquid may exist in such a correlated
electron system of the intermediate coupling region, of which there exist
anyonic excitations.

\textit{The topological Hubbard model on honeycomb lattice }: The
Hamiltonian of the topological Hubbard model on honeycomb lattice is given
by
\begin{equation}
H=H_{\mathrm{H}}+U\sum\limits_{i}\hat{n}_{i\uparrow}\hat{n}%
_{i\downarrow}-\mu\sum\limits_{i,\sigma}\hat{c}_{i\sigma}^{\dagger}\hat{c}%
_{i\sigma}.  \label{model}
\end{equation}
Here $H_{\mathrm{H}}$ is the spinful Haldane model as%
\begin{equation*}
H_{\mathrm{H}}=-t\sum\limits_{\left\langle {i,j}\right\rangle ,\sigma}\left(
\hat{c}_{i\sigma}^{\dagger}\hat{c}_{j\sigma}+h.c.\right)
-t^{\prime}\sum\limits_{\left\langle \left\langle {i,j}\right\rangle
\right\rangle ,\sigma}e^{i\varphi_{ij}}\hat{c}_{i\sigma}^{\dagger}\hat{c}%
_{j\sigma}.
\end{equation*}
$t$ and $t^{\prime}$ are the real hopping between the first-neighbor and the
second-neighbor on the different and the same sublattices, respectively. $%
e^{i\varphi_{ij}}$ is a complex phase into the second-neighbor hopping, and
we set the direction of the positive phase is clockwise $\left( \left\vert
\varphi_{ij}\right\vert =\frac{\pi}{2}\right) $\cite{Haldane}. $U$ is the
on-site Coulomb repulsion. $\sigma$ are the spin-indices representing
spin-up $(\sigma=\uparrow)$ and spin-down $(\sigma=\downarrow)$ for
electrons. $\mu$ is the chemical potential and $\mu=U/2$ when the system is
half-filling (in this paper we only study the case of half-filling). $%
\left\langle {i,j}\right\rangle $ and $\left\langle \left\langle {i,j}%
\right\rangle \right\rangle $ denote two sites on a first-neighbor and a
second-neighbor link, respectively. $\hat{n}_{i\uparrow}$ and $\hat{n}%
_{i\downarrow}$ are the number operators of electrons with up-spin and
down-spin respectively. It is obvious that there is spin rotation symmetry
but no time reversal symmetry for the topological Hubbard model.

For free fermions (the on-site Coulomb repulsion $U$ is zero), the spectrum
is $\mathbf{E}_{\mathbf{k}}=\pm\sqrt{\left( \xi_{\mathbf{k}}\right)
^{2}+\left( \xi_{\mathbf{k}}^{\prime}\right) ^{2}}$ where $\left\vert \xi_{%
\mathbf{k}}\right\vert =t\sqrt{3+2\cos{(\sqrt{3}k_{y})}+4\cos{(3k_{x}/2)}\cos%
{(\sqrt{3}k_{y}/2)}}$ and $\xi_{\mathbf{k}}^{\prime}=2t^{\prime}\sum%
\limits_{i}\sin{(\mathbf{k}\cdot\mathbf{b}_{i})}$. The parameters $\mathbf{%
a_{1},a_{2}}$ and $\mathbf{a_{3}}$ are the displacement from one site to its
nearest neighborhoods and $\mathbf{b_{1}}=\mathbf{a_{2}}-\mathbf{a_{3}}$, $%
\mathbf{b_{2}}=\mathbf{a_{3}}-\mathbf{a_{1}}$, etc. The length of the
hexagon side has been chosen to be unit. One can see that there exist an
energy gap $\Delta_{c}=6\sqrt{3}t^{\prime}$ at the points $\mathbf{k}_{1}=%
\frac{2\pi}{3}(1,$ $\frac{1}{\sqrt{3}})$ and $\mathbf{k}_{2}=\frac{2\pi }{3}%
(-1,$ $-\frac{1}{\sqrt{3}})$. Due to the existence of nonzero TKNN number%
\cite{thou}, there exists the integer quantum Hall effect $\sigma _{xy}=%
\frac{2e^{2}}{h}.$ Here the parameter $2$ comes from the contributions of
electrons of up-spin and down-spin. Therefore, for the free fermions, the
ground state is a topological insulator with quantized anomalous Hall effect%
\cite{Haldane,kane,2}.

An issue is whether the topological insulator is stable for the interaction
case. To examine stability of the topological insulator against on-site
interaction, we will use the slave-rotor approach to study the topological
Hubbard model. Slave-rotor approach has been widely applied to study the
quantum liquid states near Mott transition of correlated electron systems%
\cite{rotor,Lee,Hermele,qsh,qsh1,e}. By the slave-rotor approach, we find
that chiral spin liquid appears of the intermediate coupling region.

\textit{Slave-rotor approach} : In slave-rotor approach, we represent
electronic operator $\hat{c}_{j\sigma }$ into $\hat{c}_{j\sigma }=e^{i\theta
_{j}}\hat{f}_{j\sigma }$ where the fermion spinon $\hat{f}_{j\sigma }$
represents the spin degree of freedom and $e^{i\theta _{j}}$ represents the
charge degree of freedom together with slave-rotor's constraint $%
\sum\limits_{\sigma }\hat{f}_{j\sigma }^{\dagger }\hat{f}_{j\sigma }+{\Large %
L}_{j}=1.$ Here we introduce an additional variable - the angular momentum $%
L=i\hbar \partial _{\theta }$ associated with a quantum $O(2)$ rotor $\theta
.$ Then the Hamiltonian in Eq.(\ref{model}) turns into%
\begin{align}
H_{eff}& =-t\sum\limits_{\left\langle {i,j}\right\rangle ,\sigma }(\hat{f}%
_{i\sigma }^{\dagger }\hat{f}_{j\sigma }X_{i}^{\dagger }X_{j}+h.c.)-\mu
\sum\limits_{i,\sigma }\hat{f}_{i\sigma }^{\dagger }\hat{f}_{i\sigma }
\label{eff} \\
& -t^{\prime }\sum\limits_{\left\langle \left\langle {i,j}\right\rangle
\right\rangle ,\sigma }e^{i\varphi _{ij}}\hat{f}_{i\sigma }^{\dagger }\hat{f}%
_{j\sigma }X_{i}^{\dagger }X_{j}+\frac{U}{2}\sum\limits_{i}{\Large L}_{i}^{2}
\notag \\
& +\sum\limits_{i}h_{i}(\sum\limits_{\sigma }\hat{f}_{i\sigma }^{\dagger }%
\hat{f}_{i\sigma }+{\Large L}_{i}-1)+\sum\limits_{i}\rho _{i}(\left\vert
X_{i}\right\vert ^{2}-1).  \notag
\end{align}%
where $h_{i}$ is a Lagrange multiplier for slave-rotor's constraint. $\rho
_{i}$ is a complex Lagrange multiplier for $\left\vert X_{i}\right\vert
^{2}=1$ ($X_{i}=e^{i\theta _{i}}$).

\begin{figure}[ptb]
\includegraphics[width=0.43\textwidth]{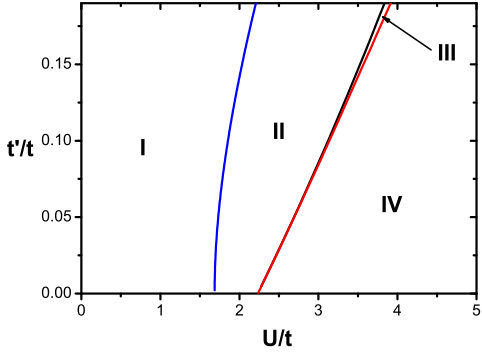}
\caption{(color online) Phase diagram at $T=0$ . There are four regions : I
is TI state, II is the quantum spin liquid, III is an AF order with QAH
effect, IV is the trivial AF order. The blue line and black line are $(\frac{%
U}{t})_{c1}$ and $(\frac{U}{t})_{c2},$ respectively.}
\end{figure}

\begin{figure}[ptb]
\includegraphics[width=0.4\textwidth]{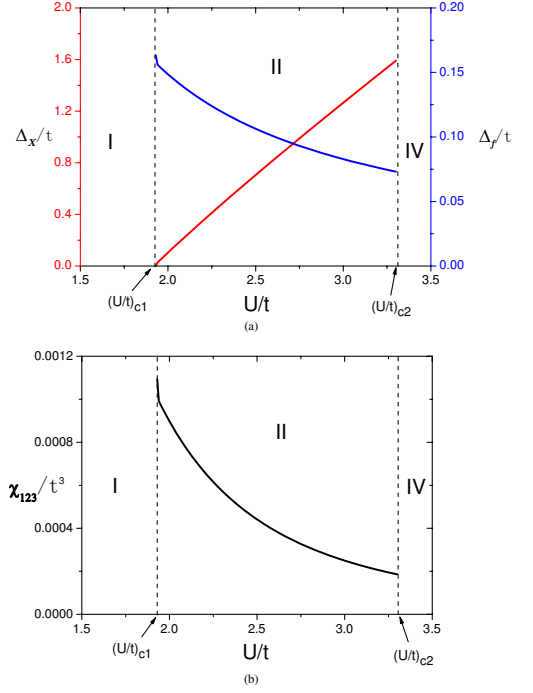}
\caption{(color online) (a) : The energy gaps of rotor $\Delta_{X}$ (the red
line) and fermionic spinon $\Delta_{f}$ (the blue line) of the topological
Hubbard model at $T=0$ and $t^{\prime}=0.12$. $\left( U/t\right) _{c1}$ and $%
\left( U/t\right) _{c2}$ are the quantum phase transitions. (b) The non-zero
spin chiral order parameter.}
\end{figure}

We introduce four variational parameters $Q_{f}=\left\langle X_{i}^{\dagger
}X_{j}\right\rangle _{ij\left\langle nn\right\rangle }$,\ $%
Q_{X}=\left\langle \sum\limits_{\sigma}\hat{f}_{i\sigma}^{\dagger}\hat{f}%
_{j\sigma}\right\rangle _{ij\left\langle nn\right\rangle }$, $%
Q_{f}^{\prime}=\left\langle X_{i}^{\dagger}X_{j}\right\rangle
_{ij\left\langle \left\langle nn\right\rangle \right\rangle }$\ and $%
Q_{X}^{\prime}=\left\langle \sum\limits_{\sigma}e^{i\varphi_{ij}}\hat{f}%
_{i\sigma}^{\dagger}\hat {f}_{j\sigma}\right\rangle _{ij\left\langle
\left\langle nn\right\rangle \right\rangle }$. To obtain the five parameters
$Q_{X},$ $Q_{f},$ $Q_{X}^{\prime},$ $Q_{f}^{\prime},$ $\rho,$ we solve the
following equations self-consistently,
\begin{align}
Q_{X} & =\frac{1}{3tN_{s}}\sum\limits_{k}\frac{Q_{f}\left\vert \xi _{\mathbf{%
k}}\right\vert ^{2}}{E_{f}},\text{ }Q_{X}^{\prime}=\frac {1}{%
3t^{^{\prime}}N_{s}}\sum\limits_{k}\frac{Q_{f}^{\prime}\left\vert \xi_{%
\mathbf{k}}^{\prime}\right\vert ^{2}}{E_{f}}, \\
Q_{f} & =\frac{1}{N_{s}}\sum\limits_{k}\frac{\left\vert \xi_{\mathbf{k}%
}\right\vert }{12t}\frac{U}{\sqrt{U\left( \rho+\varepsilon_{k}\right) }},%
\text{ }1=\frac{1}{N_{s}}\sum\limits_{k}\frac{U}{2\sqrt{U\left(
\rho+\varepsilon_{k}\right) }},  \notag \\
Q_{f}^{\prime} & =\frac{1}{N_{s}}\sum\limits_{k}\frac{g_{\mathbf{k}}}{24}%
\frac{U}{\sqrt{U\left( \rho+\varepsilon_{k}\right) }},  \notag
\end{align}
where
\begin{equation}
g_{\mathbf{k}}=4\cos\left( 3k_{x}/2\right) \cos(\sqrt{3}k_{y}/2)+2\cos (%
\sqrt{3}k_{y})
\end{equation}
and $\varepsilon_{k}=-Q_{X}\left\vert \xi_{\mathbf{k}}\right\vert -t^{\prime
}Q_{X}^{\prime}g_{\mathbf{k}}$. $N_{s}$ denoting the number of unit cells.

After the calculation, we find that the topological insulator is stable
below a critical interaction strength, $\frac{U}{t}<(\frac{U}{t})_{c1}$ (See
the results in Fig.1) With increasing interaction strength, we get non-zero
solutions of $Q_{X},$ $Q_{f},$ $Q_{X}^{\prime},$ $Q_{f}^{\prime},$ $\rho.$
As a result, the ground state turns into a quantum spin liquid state
characterized by a finite gap of rotor excitation. The excitations are not
fermions generated by $\hat{c}_{i}^{\dagger},$ instead, they are rotors and
fermionic spinons.

Then we get the mean field effective Hamiltonian as $H_{eff}=H_{f}+H_{X}$
where
\begin{equation}
H_{f}=-t\sum\limits_{\left\langle {i,j}\right\rangle ,\sigma}Q_{f}\hat {f}%
_{i\sigma}^{\dagger}\hat{f}_{j\sigma}-t^{\prime}\sum\limits_{\left\langle
\left\langle {i,j}\right\rangle \right\rangle
,\sigma}Q_{f}^{^{\prime}}e^{i\varphi_{ij}}\hat{f}_{i\sigma}^{\dagger}\hat{f}%
_{j\sigma}+h.c.  \label{hf}
\end{equation}
and
\begin{align}
H_{X} & =-t\sum\limits_{\left\langle {i,j}\right\rangle }Q_{X}X_{i}^{\ast
}X_{j}-t^{\prime}\sum\limits_{\left\langle \left\langle {i,j}\right\rangle
\right\rangle }Q_{X}^{\prime}X_{i}^{\ast}X_{j}+\frac{U}{2}\sum\limits_{i}%
{\Large L}_{i}^{2} \\
& +h\sum\limits_{i}{\Large L}_{i}+\rho\sum\limits_{i}\left\vert
X_{i}\right\vert ^{2}+h.c.  \notag
\end{align}
After diagonalizing the Hamiltonian in Eq.(\ref{hf}), we get the energy
spectrum for the two-flavor fermionic spinons $E_{f}=\pm\sqrt{%
Q_{f}^{2}\left\vert \xi_{\mathbf{k}}\right\vert
^{2}+Q_{f}^{\prime2}\left\vert \xi_{\mathbf{k}}^{\prime}\right\vert ^{2}}.$
One may get the energy gap for the fermionic spinons as $\Delta_{f}=6\sqrt{3}%
t^{\prime}Q_{f}^{\prime}$. On the other hand, the spectrum of the charge
excitations $E_{X}$ is $E_{X}=2\sqrt{U\rho+U\varepsilon_{\mathbf{k}}}.$ In
the quantum spin liquid state, $\Delta_{X}$ is not zero, $\Delta_{X}=2\sqrt{%
U\left( \rho+\min\left( \varepsilon_{\mathbf{k}}\right) \right) };$ when
approaching the phase transition between topological insulator and quantum
spin liquid state it becomes zero due to rotor condensation. From Fig.2.(a),
we can notice that the rotor's gap is much larger than the fermionic
spinon's gap (the energy scale of $\Delta_{X}$ is $10$ times to that of $%
\Delta_{f}$).

By this method we get a quantum spin liquid state with fermionic spinons by
adding the on-site interaction to the spinful Haldane model. When further
increasing the interaction strength, the quantum spin liquid is unstable
against antiferromagnetic (AF) spin density wave (SDW) order. Such AF-SDW
order is described by $\langle\hat{c}_{i,\sigma}^{\dagger}\hat{c}_{i,\sigma
}\rangle=\frac{1}{2}(1+(-1)^{i}\sigma M).$ Here $M$ is the staggered
magnetization. In HF\ mean field approach, we may get the self-consistency
equation for $M$ by minimizing the ground state energy. To characterize
different orders of the topological Hubbard model (the topological
insulator, the quantum spin liquid, the AF-SDW), we plot a phase diagram in
Fig.1. $(\frac{U}{t})_{c2}$ denotes another critical interaction strength
that divides the quantum spin liquid and the AF-SDW (See Fig.1). In
particular, in Fig.1, one may see that there exists a narrow window between
quantum spin liquid state and the trivial AF order - an AF-SDW order with
quantized anomalous Hall effect.

\textit{Effective Chern-Simons theory }: In the following parts, we will
focus on the quantum spin liquids between topological insulator and AF-SDW
state (region II\ in Fig.1).

In the quantum spin liquid state, because the rotor excitation $X_{i}$ has a
big energy gap, we may integrate out it and concentrate only on the spinon
excitations. The fluctuations of $h_{i}$ and the phase fluctuations of $%
Q_{f},$ $Q_{f}^{/}$ amount to coupling the fermionic spinons to a compact
\textrm{U(1)} gauge field $a_{ij}$ by the minimal prescription. After
considering the fluctuations around the mean field saddle point, we get the
effective model of fermionic spinons with \textrm{U(1)} gauge invariance
\begin{eqnarray}
L_{f} &=&\sum_{j,\sigma }\hat{f}_{j\sigma }^{\ast }(\partial _{\tau
}-ia_{\tau ,j}+h_{0}-\mu )\hat{f}_{j\sigma } \\
&&-tQ_{f}\sum\limits_{\left\langle {i,j}\right\rangle ,\sigma }e^{ia_{ij}}%
\hat{f}_{i\sigma }^{\dagger }\hat{f}_{j\sigma }  \notag \\
&&-t^{\prime }Q_{f}^{\prime }\sum\limits_{\left\langle \left\langle {i,j}%
\right\rangle \right\rangle ,\sigma }e^{i\varphi _{ij}}e^{ia_{ij}}\hat{f}%
_{i\sigma }^{\dagger }\hat{f}_{j\sigma }+h.c.  \notag
\end{eqnarray}%
where $h_{0}=\left\langle h_{i}\right\rangle =\mu $. Hence the continuum
version of above model becomes the two flavor massive Schwinger model with
the Lagrangian as $\mathcal{L}_{\mathrm{f}}=i\bar{\psi}\gamma _{\mu
}(\partial _{\mu }-ia_{\mu })\psi +m\bar{\psi}\psi $ where $m=\Delta _{f}/2$
is a fermion mass and $\bar{\psi}_{\alpha }=\psi _{\alpha }^{\dagger }\gamma
_{0}=(%
\begin{array}{llll}
\bar{f}_{\uparrow \alpha A}, & \bar{f}_{\uparrow \alpha B}, & \bar{f}%
_{\downarrow \alpha A}, & \bar{f}_{\downarrow \alpha B}%
\end{array}%
)$ and $\alpha =1,$ $2$ labels the two points $\mathbf{k}_{1}=\frac{2\pi }{3}%
(1,$ $\frac{1}{\sqrt{3}})$ and $\mathbf{k}_{2}=\frac{2\pi }{3}(-1,$ $-\frac{1%
}{\sqrt{3}})$.

Considering the quantum fluctuations of fermionic spinons, we get a two
dimensional dynamics Maxwell model of the gauge field $a_{\mu}$ as $\mathcal{%
L}_{a}=\frac{1}{4e_{a}^{2}}(\partial_{\mu}a_{\nu})^{2}$ where $e_{a}^{2}=%
\frac{1}{3\pi}\frac{1}{m}.$ The compact \textrm{U(1)} gauge theory is always
confining\cite{confine}. However, the induced Chern-Simons (CS) term will
lead to deconfinement. Integrate over fermions by using $1/m$ (gradient)
expansion approach, we obtain the CS term $\mathcal{L}_{cs}=\frac{N}{2}\frac{%
m}{|m|}\frac{1}{4\pi}\epsilon^{\mu\nu\lambda}a_{\mu}\partial_{\nu
}a_{\lambda}$ where $N=4$ (two-flavor plus two spin components)\cite%
{redlich,cs}.

Finally we obtain an effective CS theory of the quantum spin liquid state
with the Lagrangian $\mathcal{L}_{\mathrm{eff}}=\mathcal{L}_{\mathrm{f}}+%
\mathcal{L}_{a}+\mathcal{L}_{CS}.$

\textit{Chiral spin liquid} : After obtaining the effective CS\ theory, the
quantum spin liquid state ( region II in Fig.1) is identified to chiral spin
liquid. Such a topologically ordered spin liquid breaks time-reversal
symmetry while preserves all other symmetries (spin rotation symmetry,
translation symmetry, ...).

Firstly we point out that the quasi-particle is really anyon. Let us
consider single $\pi $-flux excitation ($\Phi =\pi $) in the quantum spin
liquid as shown in Fig.3. The vacuum expectation value of the fermion number
$\langle N^{\mathrm{f}}\rangle $ is related to the spectral asymmetry of the
Dirac Hamiltonian
\begin{equation}
\langle N^{\mathrm{f}}\rangle =-\frac{1}{2}\int_{-\infty }^{\infty }dE\,%
\frac{1}{\pi }\text{\textrm{Im} }\mathrm{Tr\,}(\frac{1}{H_{f}-E-i\epsilon }%
)\,\,\text{\textrm{sign}}(E)
\end{equation}%
where $H_{f}$ is the Hamiltonian of the fermion spinon\cite%
{NiemiSemenoff-1986}. And the fermionic number is also related to the
Atiah-Patodi-Singer invariant $\mathcal{\eta }_{_{H}}=-\frac{1}{2}\langle N^{%
\mathrm{f}}\rangle $ which represents the difference between the number of
states with positive and negative energy. The Atiah-Patodi-Singer index
theorem states that due to the quantum anomaly the fermionic number of the
Dirac operator, equals the topological charge as $\langle N^{\mathrm{f}%
}\rangle =-\frac{N}{2}\frac{m}{|m|}\frac{\Phi }{2\pi }$, $N=4.$ It
represents a fact that a $\pi $-flux excitation with half topological charge
carries one fermion number, $\left\vert \langle N^{\mathrm{f}}\rangle
\right\vert =1.$ That means a $\pi $-flux excitation is really a bound state
of $\pi $-flux and a fermionic spinon. Due to nontrivial AB phases upon
adiabatic exchange of charge and flux, $\pi $-flux turns into semion -
special type of Abelian anyon. By the fusion rules of Abelian anyons, one
may find a statistical angle $\theta $ is $\frac{1}{2}=\frac{\pi }{2}$.

Secondly, we calculate the topological degeneracy of CSL, a topologically
ordered spin liquid. In the temporal gauge, $a_{0}=0,$ and on a torus, the
ground states are characterized by zero momentum gauge fields $(a_{x,\mathbf{%
k=0}},$ $a_{y,\mathbf{k=0}})$. After straightforwardly calculations\cite{7},
we may get the effective Hamiltonian of $(a_{x,\mathbf{k=0}},$ $a_{y,\mathbf{%
k=0}})$ as $\mathcal{H}_{eff}=\frac{(\mathcal{P}_{\theta_{x}}-\mathcal{A}%
_{\theta_{y}})^{2}}{2M_{x}}+\frac{(\mathcal{P}_{\theta_{y}}-\mathcal{A}%
_{\theta_{x}})^{2}}{2M_{y}}$ where $\mathcal{A}_{\theta_{x}}=-\frac{a_{y,%
\mathbf{k=0}}}{2\pi}$, $\mathcal{A}_{\theta_{y}}=\frac{a_{x,\mathbf{k=0}}}{%
2\pi}$ and $M_{x}=\frac{1}{e_{a}^{2}}\frac{L_{y}}{L_{x}},$ $M_{y}=\frac{1}{%
e_{a}^{2}}\frac{L_{x}}{L_{y}}$ ($L_{x}$ and $L_{y}$ are the lengths of the
system along x- and y-directions, respectively). This model corresponds to a
particle on a plane with a finite "magnetic field". The strength of the
"effective magnetic field" is\textbf{\ }obtained as $\mathcal{B}_{eff}=\frac{%
2}{2\pi}.$ There exists a two-unit flux tube through the center of the
torus. So the degeneracy is given as $\mathcal{D}=2$.

Next we calculate the edge states of the CSL from the effective CS theory.
We know that the charges of $a_{\mu}$ are quantized as integers. Then the
effective CS theory has two right-moving edge excitations. The two branches
of the edge excitations are described by the following 1D fermion theory$%
\mathcal{L}_{\text{\textrm{edge}}}=\sum_{\alpha}\psi_{\alpha R}^{\dag
}(\partial_{t}-v_{R}\partial_{x})\psi_{\alpha R},$ where $\alpha=1,2.$ $%
\psi_{\alpha R}$ carries a unit of $a_{\mu}$ charge. That means we get
spin-charge separated edge states : the edge modes carry only spin current%
\cite{edge0,edge}.

\begin{figure}[ptb]
\includegraphics[width=0.4\textwidth]{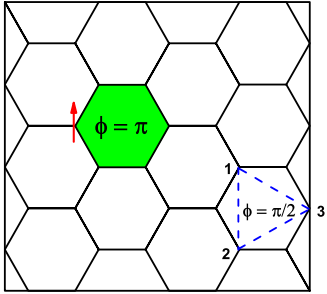}
\caption{(color online) Scheme of an anyon as a composite object of a
fermionic spinon and a $\protect\pi$-flux. $1,$ $2,$ $3$ correspond three
vertices of a equilateral triangle to define the spin chiral order parameter
in Eq.(\protect\ref{chiral}).}
\end{figure}

Thirdly, an important property of CSL is the non-zero spin chiral order
parameter which has a non zero expectation value in a phase with broken $P$
and $T$ symmetry. Spin chiral order parameter is\textit{\ }a rotationally
invariant operator defined through\cite{chiral,ln}
\begin{align}
\mathcal{\chi}_{_{\langle123\rangle}} & =\left\langle {\mathbf{S}}_{1}\cdot({%
\mathbf{S}}_{2}\times{\mathbf{S}}_{3})\right\rangle  \label{chiral} \\
& =\frac{1}{4i}\left\langle \hat{f}_{1\alpha}^{\dagger}\hat{f}_{2\alpha}\hat{%
f}_{2\beta}^{\dagger}\hat{f}_{3\beta}\hat{f}_{3\gamma}^{\dagger}\hat {f}%
_{1\gamma}-\hat{f}_{1\alpha}^{\dagger}\hat{f}_{3\alpha}\hat{f}_{3\beta
}^{\dagger}\hat{f}_{2\beta}\hat{f}_{2\gamma}^{\dagger}\hat{f}_{1\gamma
}\right\rangle .  \notag
\end{align}
If the sites $1,$ $2,$ $3$ correspond three vertices of a equilateral
triangle in a plaquette, we may estimate the mean field value of $\mathcal{%
\chi }_{_{\langle123\rangle}}$ to be $\frac{1}{2}\left( \sin\Phi\right)
\left\vert Q_{X}^{\prime}\right\vert ^{3}$ where $\Phi$ is the gauge
invariant flux through the equilateral triangle, $\Phi=\left\vert
\varphi_{12}+\varphi_{23}+\varphi_{31}\right\vert =\frac{\pi}{2}$. Thus we
get non-zero spin chiral order parameter along these loops. See the results
in Fig.2.(b).

In summary. we have predicted an emergent chiral spin liquid state base on
the topological Hubbard model on honeycomb lattice with spin rotation
symmetry. In the end, we address the relevant experimental realization and
the way to be conformed by numerical approaches. In condensed matter
physics, there is no such material with a Hamiltonian of the topological
Hubbard model. However, such system may be simulated in optical lattice of
cold atoms. In Ref.\cite{wu,zhu}, it is proposed that the (spinless) Haldane
model on honeycomb optical lattice can be realized in the cold atoms. When
two-component fermions with repulsive interaction are put into such optical
lattice, one can get an effective topological Hubbard model. It is easy to
change the potential barrier by varying the laser intensities to tune the
Hamiltonian parameters including the hopping strength ($t$-term) and the
particle interaction ($U$-term). On the other hand, one may check our
prediction by quantum Monte Carlo (QMC) simulations including the global
phase diagram, the topological degeneracy, the spin chiral order parameter%
\cite{chiral}.

The authors acknowledge that this research is supported by SRFDP,
NFSC Grant No. 10874017 and 10774015, National Basic Research
Program of China (973 Program) under the grant No. 2011CB921803,
2011cba00102.

\end{document}